\documentstyle[12pt]{article}
\begin{document}
\begin{center}
{\Large\bf LEP, SLC AND THE STANDARD MODEL}
\vspace{1cm}

{ {\large\bf{V.NOVIKOV,  L.OKUN, M.VYSOTSKY}} \\
ITEP, Moscow, 117259, Russia \\

\vspace{1cm}
{ {\large\bf A.ROZANOV}} \\
ITEP and CPPM-IN2P3-CNRS, Marseille, France}
\end{center}
\date{}

\begin{abstract}
 A simple way of deriving and analyzing electroweak radiative
corrections to the $Z$ boson decays is presented in the framework
of the Standard Model.  This paper is an updated and revised
version of a plenary talk given at the Second Sakharov
Conference; Lebedev Physical Institute, Moscow, May 20-24, 1996.
\\
\end{abstract}
\section{LEP-I and SLC}
LEP-I (CERN) and SLC (SLAC) electron-positron colliders had
started their operation in the fall 1989.
The  sum of energies of $e^+ + e^-$ was
chosen to be equal to the $Z$ boson mass.  LEP-I was terminated in the
fall of 1995  in order to give place to LEP-II which operated
at energy 135 GeV in spring 1996, 172 GeV in spring 1997 and will
finally reach about 200 GeV in 1999 -- 2000. Meanwhile SLC continues
at energy close to 91 GeV.
The reactions, which have been  studied at LEP-I (detectors: ALEPH,
DELPHI, L3, OPAL) and SLC (detector SLD) may be  presented in the form:
$$e^+e^- \to  Z \to f\bar{f}\;\;,$$
where
\begin{center}
\vspace{3mm}
\begin{tabular}{ll}
$f\bar{f}$ = &
$\nu\bar{\nu}(\nu_e\bar{\nu}_e~,~~\nu_{\mu}\bar{\nu}_{\mu}~,
~~\nu_{\tau}\bar{\nu}_{\tau})$ -- invisible~, \\
      & $l\bar{l}(e\bar{e}~, ~~\mu\bar{\mu}~, ~~\tau\bar{\tau})$
      -- charged leptons~,\\
& $q\bar{q}(u\bar{u}~,~d\bar{d}~,~s\bar{s}~,~c\bar{c}~,
b\bar{b}) \to $  hadrons~.
\end{tabular}
\end{center}
\vspace{3mm}
About 20,000,000 $Z$ bosons
has been detected at LEP and about 100,000 at SLC (but here electrons
are polarized). Fantastic precision has been reached in the
measurement of the $Z$ boson properties \cite{1,2,3}:
$$M_Z = 91,186.3 \pm 1.9
\mbox{\rm MeV}\;, \;\; \Gamma_Z = 2,494.7 \pm 2.6 \mbox{\rm MeV}\;,$$
$$\Gamma_h \equiv \Gamma_{hadrons} = 1,743.6 \pm 2.5 \mbox{\rm MeV}\;,
\;\; R_l = \Gamma_h/\Gamma_l = 20.783 \pm 0.029 \;,$$
$$\Gamma_{invisible} = 499.5 \pm 2.0 \mbox{\rm MeV}\;,$$
where $\Gamma_l$ refers to the width of a single charged leptonic
channel. (As a rule we use the most recent data submitted to the
XXIInd Rencontre de Moriond Conference, Les Arcs/Savoie, France,
March 15-22, 1997 \cite{1}. These data have tiny deviations from the
data reported at the 28th ICHEP, Warsaw, Poland, 25-31 July, 1996
\cite{2}, or from the data from the latest Review of Particle
Physics \cite{3}.)

By comparing the $\Gamma_{invisible}$ with  theoretical predictions
for neutrino decays it was established that the number of neutrinos
which interact with $Z$ boson is 3:  ($N_{\nu} = 2.989 \pm 0.012$).
This is a result of fundamental importance.

More than 2,000 experimentalists and engineers and hundreds of
theorists participated in this unique collective quest for truth.

\section{Theoretical analysis: the fundamental parameters.}

The theoretical study of electroweak radiative corrections started in
1970's and was elaborated by a number of theoretical groups in many
countries \cite{3} - \cite{221}. This study has been summarized in
Yellow CERN Reports of working groups on precision calculations for
the $Z$ resonance in 1995 \cite{1144}. The results of different groups
are in good agreement. The deviations in theoretical calculations are
by order of magnitude smaller than the experimental uncertainties.
In what follows we mainly use our own
calculations \cite{221} in which we tried to simplify the work of
previous authors \cite{3}-\cite{14} by separating genuine electroweak
radiative corrections from purely electromagnetic ones.

It is instructive to compare the electroweak theory with the quantum
electrodynamics (QED). In the latter there are two fundamental
parameters: mass of the electron, $m_e$, and its charge, $e$, or fine
structure constant $\alpha = e^2/4\pi$ (we use units $\hbar, c =1$).
Both are not only fundamental, but also known with high precision.
Every observable in QED can be expressed in terms of $m_e$ and
$\alpha$ (and, of course, of energies and momenta of particles
participating in a given process).

In the electroweak theory the situation is more complex for several
reasons:
\begin{enumerate}
\item{There are more fundamental charges and masses.}
\item{They are not independent of each other.}
\item{Not all of them are known with high accuracy.}
\item{On the other hand, as a basic parameter of the theory a
quantity is used, which is
known with highest accuracy, but which is not fundamental, the
four-fermion coupling of the muon decay, $G_{\mu}$.}
\end{enumerate}

The fundamental masses of the electroweak theory are masses of $W$ and
$Z$ bosons, $m_W$ and $m_Z$. Among the masses of fermions, the most
important for the $Z$-decay  is the mass of the top-quark, $m_t$.

The fundamental couplings of the electroweak theory are $e,~f,~g$, or
$\alpha = e^2/4\pi$, $\alpha_Z = f^2/4\pi$, $\alpha_W = g^2/4\pi$:
$e$ is the coupling of photons to electrically charged particles,
$f$ is the coupling of $Z$ bosons to weak neutral current, e.g.
$\bar{\nu}\nu$,
$g$ is the coupling of $W$ bosons to weak charged current, e.g.
$\bar{e}\nu$.

While the charged current is a purely V-A current of the form
$\gamma_{\alpha}(1+\gamma_5)$, the ratio  $g_{Vf}/g_{Af}$ between the
vector and axial vector neutral currents depends on the third
projection of the isotopic spin of the fermion $f$, $T^f_3$, and on
its electric charge $Q^f$. The decay amplitude of the $Z$ boson
may be written in the form:
$$M(Z \to f\bar{f}) = \frac{1}{2}f
\bar{\psi}_f (g_{Vf} \gamma_{\alpha} + g_{Af}
\gamma_{\alpha}\gamma_5) \psi_f Z^{\alpha}\;\;,$$
where $\bar{\psi}_f$ -- is the wave function of emitted fermion, $\psi_f$
corresponds to the emitted antifermion (or absorbed fermion),
$Z^{\alpha}$ is the wave function of the $Z$ boson. At the tree level
(see, e.g. text-books \cite{144}):
$$ g_{Af} = T_3^f\;\;, \;\; g_{Vf} = T_3^f - 2Q^f s^2_W\;.  $$
Here $$
T_3^f = +1/2 \;\; \mbox{\rm for} \;\; f = \nu,~ u,~ c\;; \;\;
T^f_3 = -1/2 \;\; {\rm for} \;\; f = l,~ d,~ s,~ b~.$$
Thus $$g_{Vf}/g_{Af} = 1 - 4|Q^f| s^2_W\;\;.$$
In the above expressions $s_W \equiv \sin \theta_W$, where $\theta_W$
-- is the so called weak angle. At the tree level (no loops):
$e/g = s_W\;\;, \;\;g/f = c_W\;\;, \;\; m_W/m_Z = c_W\;\;,\;\;$
 where $c_W \equiv \cos \theta_W$.

The four-fermion coupling constant $G_{\mu}$ is extracted from the
life-time of the muon, $\tau_{\mu}$, after taking into account
the well-known kinematic and electromagnetic corrections:

$$1/\tau_{\mu} = \Gamma_{\mu} = \frac{G^2_{\mu} m^5_{\mu}}{192 \pi^3}
(1 + \mbox{\rm well~~ known~~ corrections}\;
\sim (\frac{m_e}{m_{\mu}})^2 ,\;
\alpha)\;\;,$$

$$G_{\mu} = (1.16639 \pm 0.00002) \cdot 10^{-5} \mbox{\rm GeV}^{-2}\;.$$

In the tree approximation the four-fermion coupling constant
$G_{\mu}$ can be expressed in terms of $W$ boson coupling  constant
$g$ and its mass $m_W$:

$$G_{\mu} = \frac{g^2}{4\sqrt{2}m^2_W} = \frac{\pi\alpha}{\sqrt{2}
m^2_W s^2_W} = \frac{\pi\alpha}{\sqrt{2} m^2_Z s^2_W c^2_W} \;.$$

(The last two expressions are derived by using the relations
$e/g = s_W$, $\alpha = e^2/4\pi$, $m_W/m_Z = c_W$).

\section{Theoretical analysis: the running $\alpha(q^2)$.}

It has been well known since 1950's that  electric charge $e$ and
hence $\alpha$ logarithmically depend on the square of the
four-momentum of the photon, $q^2$ \cite{16}. (For a real photon $q^2= 0$,
for a virtual one $q^2 \neq 0$). This phenomenon is usually
referred to as "the running of $\alpha$". It is caused by vacuum
polarization, by loops of virtual charged particles: charged leptons,
$l\bar{l}$, and quarks, $q\bar{q}$, inserted into the propagator of a
photon. As a result $\bar{\alpha} \equiv \alpha(q^2 = m^2_Z)$ is
approximately by 6\% larger than $\alpha \equiv \alpha(0)$.

The relation between $\bar{\alpha}$ and $\alpha$ is obtained by
summing up an infinite chain of loops:

$\bar{\alpha} = \alpha/(1 - \delta\alpha)$;

$\delta\alpha = \delta\alpha_l + \delta\alpha_h$, where
$\delta\alpha_l$ is the one-loop contribution of three
charged leptons, while $\delta\alpha_h$ -- is that of
 five quarks (hadrons). The leptonic contribution is predicted
 with very high accuracy:  $\delta\alpha_l =
 \frac{\alpha}{3\pi}\sum_{l}(\ln\frac{m_Z^2}{m_l^2}-\frac{5}{3}) =0.03141$.

 The hadronic contribution is obtained on 
 the basis of dispersion relations and low-energy experimental data
 on $e^+e^-$-annihilation into hadrons \cite{17}:
$\delta\alpha_h = 0.02799(66)$.

The value of $\alpha(0)$ is known with extremely high accuracy:\\
$\alpha \equiv \alpha(0) = 1/137.035985(61)$;
the accuracy of
$\alpha$ is very important for QED, but irrelevant to electroweak
physics.

The value of $\bar{\alpha}$ is less accurate:
$\bar{\alpha} = 1/128.896(90)$ \cite{17}, but $\bar{\alpha}$ is
pivotal for electroweak physics. Let us stress
that the running of $\alpha(q^2)$ is a purely electromagnetic effect,
caused by electromagnetic loops of light fermions. Contributions of
$t\bar{t}$ and $W^+W^-$ are negligibly small and may be taken into
account together with purely electroweak loops.

Unlike $\alpha(q^2)$, two other electroweak couplings $\alpha_Z(q^2)$
and $\alpha_W(q^2)$ are not running but "crawling" in the interval
$0 \leq q^2 \leq m_Z^2$ \cite{18}:
$$\alpha_Z(m^2_Z) = 1/22.91, \;\; \alpha_Z(0) = 1/23.10 \;\;;$$

$$\alpha_W(m^2_Z) = 1/28.74, \;\; \alpha_W(0) = 1/29.01\;\;.$$

The natural scale for $Z$-physics is $q^2 = m^2_Z$. Therefore it is
evident that $\bar{\alpha} \equiv \alpha(m^2_Z)$, not
$\alpha \equiv \alpha(0)$ is the relevant parameter.
In fact, in all computer codes,
dealing with $Z$-physics, $\bar{\alpha}$  enters at a certain stage
and substitutes $\alpha$. But this occurs inside the "black box" of
the code, while $\alpha$ formally plays the role of an input
parameter.  In these codes the running of $\alpha$ is considered as
(the largest) electroweak correction. We consider this running as
purely electromagnetic one and define our Born approximation in terms
of $\bar{\alpha}$, $G_{\mu}$ and $m_Z$.

Instead of angle $\theta_W$, we define angle $\theta$ \cite{8,221}
($s \equiv \sin \theta\;, \;\; c \equiv \cos \theta$) in the following
way:
$$G_{\mu} = \frac{g^2(q^2 = 0)}{4\sqrt{2} m^2_W} \simeq
\frac{g^2(q^2=m_Z^2)}{4\sqrt{2} m_W^2}\;\;,$$
where the second equality is based on the "crawling" of $g(q^2)$:
$g(0) \simeq g(m^2_Z)$. We use it to define the angle
$\theta$:
$$ G_{\mu} \equiv \frac{e^2(m_Z^2)}{4\sqrt{2}s^2 m_W^2} =
 \frac{\pi\bar{\alpha}}{\sqrt{2}s^2 c^2 m_Z^2}  \;\; . $$
Thus,
$$ \sin^2  2\theta = \frac{4\pi\bar{\alpha}}{\sqrt{2} G_{\mu} m^2_Z} =
 0.71078(50)\;\;,$$
$$s^2 = 0.23110(23)\;\;,\;\; c^2 =
0.76890(23)\;\;, \;\; c = 0.87687(13)\;\;.$$

Our Born approximation
starts with the most accurately known observables: 
$ G_{\mu}\;, \;m_Z\;, \; \bar{\alpha}\;\; ({\rm or}\; s^2)$ \cite{221}.

Traditionally parametrization in terms of
$G_{\mu}\;, \; \alpha$, and Sirlin's $\theta_W$ \cite{6}
$s^2_W \equiv 1 - m^2_W/m^2_Z$ is widely used in the
literature (see e.g. the review \cite{3} and references therein).
This parametrization is less convenient ($s_W$ has poor accuracy:
$\Delta m_W = \pm 80$ MeV; running of $\alpha$ is not separated from
electroweak corrections and overshadows them). (In references
\cite{99} the Born approximation is also defined by $s^2$, but purely
electromagnetic "radiators" are included in the definitions of
electroweak observables.)

\section{Theoretical analysis: one-loop electroweak corrections.}

For the sake of brevity let us choose two observables:
$$s^2_W \equiv 1 - \frac{m^2_W}{m^2_Z} \;\;, \;\;
s^2_l \equiv \frac{1}{4}(1 - \frac{g_{Vl}}{g_{Al}}) \;\; .$$

In the Born approximation $s^2_W = s^2_l = s^2$. From UA2 and CDF
experiments measuring the mass of $W$ boson \cite{1}:
$$s^2_W = 0.2231(16)\;\;,
5\sigma \;\; \mbox{\rm away~~ from}\;\; s^2 = 0.23110(23)\;.$$

From the parity violating
asymmetries (and, hence $g_{Vl}/g_{Al}$) measured at LEP and SLC:
$$s^2_l = 0.23141(28)\;, \;\; 1\sigma \;\; \mbox{\rm away~~ from}\;\; s^2\;.$$
(Note the high experimental accuracy of $s^2_l$ compared to that of
$s^2_W$.)

In the one-loop approximation
$$s^2_l = s^2 - \frac{3}{16\pi} \frac{\bar{\alpha}}{c^2 - s^2} V_{R_l}
(m_t,\; m_H)\;\;,$$
where $c^2 - s^2 = 0.5378$; index $R_l$ stands for the ratio
$g_{Vl}/g_{Al}$, and the radiative correction depends on the masses
of the top quark and higgs.  These masses enter via loops containing
virtual top quark, or higgs.  The coefficient in front of $V_{R_l}$
is chosen in such a way that 
$V_{R_l}(m_t\;,\; m_H) \approx t \equiv (m_t/m_Z)^2$ for $t\gg 1$.
 The same asymptotic normalization is used
for radiative corrections to other electroweak observables
\cite{221}.  The good agreement, within $1 \sigma$, between
experimental value of $ s^2_l$ and its Born value
means that corresponding electroweak radiative correction is
anomalously small \cite{221}. The unexpected smallness of $V_{R_l}$
is the result of cancelation between large and positive contribution
from the $t$-quark loops and large and negative contribution from
loops of other virtual particles. This cancelation, which looks like
a conspiracy, occurs when $m_{top}$ is around 160 GeV, if higgs is
light $(m_H \leq 100 $ GeV). If higgs is heavy $(m_H = 1000$ GeV) it
occurs when $m_{top}$ is around 210 GeV.

  Thus, vanishing electroweak 
radiative corrections told us that top was heavy before it was
discovered.

\section{LEPTOP and the general fit.}

The analytical formulas for all electroweak observables have been
incorporated in our computer code which we dubbed LEPTOP \cite{19}.
The fit of all electroweak data by LEPTOP gives:
$$m_t = 181 \pm 5^{+17}_{-21}\; \mbox{\rm GeV}\;\; .$$
The central value $(181 \pm 5)$ corresponds to
$1/\bar{\alpha} = 128.896$ and $m_H = 300$ GeV; the
shifts (+17, -21) -- to 
$m_H = 1000$ and 60 GeV, respectively.
This prediction is in perfect agreement with the recent (spring 1997)
data on the direct measurements of the top mass by two collaborations
at FNAL:
$$m_t = 175.6 \pm 5.5 \; {\rm (CDF/D0)\;\; \cite{1}}\;\;.$$

Unfortunately the electroweak radiative corrections depend 
on $m_H$ only weakly (are proportional to $\ln m_H/m_Z$). 
This weak dependence results in a rather poor accuracy for $m_H$.
The values of $m_t$ and $m_H$
from four parameter fit  ($m_t, m_H, \hat{\alpha}_s, \bar{\alpha}$)
with the constraints by world average values of
$\hat{\alpha}_s^{world} =0.118 \pm 0.003$ and
 $\bar{\alpha}^{world} = 1/128.896(90)$ are:

$$m_t = 172.6 \pm 5.3 \;\; \mbox{GeV} ,$$

$$m_H = 134 ^{+119}_{-72} \;\; \mbox{GeV} ,$$

$$\log m_H = 2.13 ^{+0.28}_{-0.33} \;\; ,$$

$$\chi^2/n.d.f. = 21/15 \;\; ,$$
where all LEP/SLC, $\nu N$, $M_W$ and the direct CDF/D0 $m_{top}$
measurements are taken into account ($\log m_H$ is digital logarithm and 
$m_H$
is taken in GeV).
The error of the higgs mass does not take into the account the theoretical 
uncertainties, which were estimated in \cite{1} to be
$\delta \log m_H = 0.1$.
We see that the upper bound on $m_H$ from
the analysis of electroweak radiative corrections is not yet very strong,
about 1000 GeV at $3\sigma$ level. One should
not give too much credit to the
central value of $m_H$ from the fit.
 One way to demonstrate this is to remove
$\bar{\alpha}^{world}$ from the input data set and determine
$\bar{\alpha}$
 value only from
four-parameter ($m_t, m_H, \hat{\alpha}_s, \bar{\alpha}$) 
fit to LEP, SLC, $p\bar{p}$, $\nu N$
results:
\begin{center}
\begin{tabular}{lll}
$\alpha_s$ is not constrained &~~ & $\alpha_s$ is constrained:  \\
&~~& $\hat{\alpha}_s^{world} = 0.118 \pm 0.003$ \\
$m_t = 173.8 \pm 5.3$ GeV &~~ & $ 173.2 \pm 5.3$ GeV \\
$m_H = 336^{+623}_{-328}$ GeV &~~ & $248^{+423}_{-236}$ GeV \\
$\log m_H = 2.53^{+0.46}_{-1.66}$ &~~ & $2.39^{+0.43}_{-1.35}$ \\
$\hat{\alpha}_s = 0.121 \pm 0.004$ &~~ & $0.119 \pm 0.002$ \\
$1/\bar{\alpha} = 129.107 \pm 0.254$ &~~ & $129.050 \pm 0.262$ \\
$\chi^2/n.d.f. = 20/13$ &~~ & $21/14$
\end{tabular}
\end{center}


We see that the fit of the $Z$ boson decay gives a value of 
$\bar{\alpha}$, 
compatible with that derived from the low-energy $e^+e^-$
data, although with larger uncertainties. We also see that a slight
decrease of the central value of $\bar{\alpha}$ leads to a drastic 
increase
of the central value of $m_H$ (see also \cite{305}).
This anticorrelation follows from the expression for $s^2_l$ given at
the beginning of chapter 4 and the decrease of the function $V_{R_l}$
with increase of the $m_H$ value (similar consideration holds for $s_W^2$).

Hadronic decays of $Z$ are sensitive to the value of the gluonic
coupling $\alpha_s$:

$$\Gamma_q \equiv \Gamma(Z \to q\bar{q}) = 12\Gamma_0 [g^2_{Aq}
R_{Aq} + g^2_{Vq} R_{Vq}]\;,$$

where $q$ is a generic quark, and
$$\Gamma_0 = \frac{G_{\mu} m^3_Z}{24\sqrt{2} \pi} = 82.944(6) \;
\mbox{\rm MeV}\;\;.$$

The "radiators" $R_{Aq}$ and $R_{Vq}$ contain QCD and QED corrections
caused by the final state 
emission and exchange of gluons and photons. 
As is well known \cite{3},
in the first approximation

$R_{Vq} = R_{Aq} = 1 + \frac{\hat{\alpha}_s}{\pi}\;\;, $

where $\hat{\alpha}_s \equiv \alpha_s(m^2_Z)$.
The LEPTOP fit of all electroweak
data gives:  $ \hat{\alpha}_s = 0.122(3)^{+2}_{-1}\;\;.$ Here the
central value $(0.122\pm 0.003)$ corresponds to $m_H = 300$ GeV;
the shifts $+0.002 $ and $-0.001$ -- to $m_H = 1000$ GeV and 60 GeV,
respectively.

Let us note, that low energy processes (deep inelastic scattering,
$\Upsilon$-spec\-tro\-sco\-py) give much smaller values of
$\hat{\alpha}_s$, around 0.110, when extrapolated to $q^2 = m^2_Z$.
There are different opinions on the seriousness of this discrepancy.

Another problem  was connected with the experimental value of the
width of the decay $Z \to b\bar{b}$. Theoretically the ratio
$R_b = \Gamma_b/\Gamma_h$ is not sensitive to $\hat{\alpha}_s$, $m_t$ and
$m_H$; the theory predicts:
$R_b = 0.2154(2)^{-7}_{+7}\;\;,$
where again the central value $(0.2154 \pm 0.0002)$ corresponds to
$m_H = 300$ GeV, whilst shifted by -0.0007 at $m_H = 1000$ GeV and
by $+0.0007$ at $m_H = 60$ GeV. Up to 1996 the experimental
value was $R_b = 0.2219(17)$, which was  $4\sigma$ larger than the
theoretical prediction based on the Standard Model.

In May 1996 we ended the talk at the Sakharov Conference by saying:
"Both problems (of $\hat{\alpha}_s$ and of
$R_b$) if they exist may be solved by new physics.
But maybe, experimentalists, can also change their numbers?"
Since that time
the experimental value of $R_b$ decreased by $0.0040$ to
$0.2179(11)$; now it is only $2\sigma$ away from the theoretical
prediction.

The work was supported by grants RFBR 96-02-18010, 96-15-96578,
INTAS-RFBR 95-0567, INTAS 93-3316-Ext.

\end{document}